\documentclass{article}

\usepackage[preprint]{neurips_2025}

\usepackage[utf8]{inputenc} % allow utf-8 input
\usepackage[T1]{fontenc}    % use 8-bit T1 fonts
\usepackage{hyperref}       % hyperlinks
\usepackage{url}            % simple URL typesetting
\usepackage{booktabs}       % professional-quality tables
\usepackage{amsfonts}       % blackboard math symbols
\usepackage{nicefrac}       % compact symbols for 1/2, etc.
\usepackage{microtype}      % microtypography
\usepackage{xcolor}         % colors
\usepackage{graphicx}
\usepackage{algorithm}
\usepackage[noEnd=false]{algpseudocodex}
\usepackage{amsmath}
\usepackage{graphicx}
\usepackage{epsfig}
\usepackage{multirow}
\usepackage{setspace}
\usepackage{amssymb}
\usepackage{tcolorbox}
\AtBeginEnvironment{tcolorbox}{\footnotesize}
\usepackage{lipsum}
\usepackage{subcaption}
\usepackage{amsthm}
\usepackage{listings}
\usepackage{xcolor}
\newtheorem{definition}{Definition}

\algrenewcommand\algorithmicrequire{\textbf{Input:}}
\algrenewcommand\algorithmicensure{\textbf{Output:}}

\usepackage{natbib}
\setcitestyle{authoryear,open={(},close={)}}

\title{Multimodal Chip Physical Design Engineer Assistant}

\author{%
  Yun-Da Tsai \\
  National Taiwan University \\
  NVIDIA Research \\
  \texttt{f08946007@csie.ntu.edu.tw} \\
  \And
  Chang-Yu Chao \\
  National Taiwan University \\
  \texttt{r13922001@csie.ntu.edu.tw} \\
  \And
  Liang-Yeh Shen \\
  National Taiwan University \\
  \texttt{b10507008@csie.ntu.edu.tw} \\
  \And
  Tsung-Han Lin \\
  University of California, Los Angeles \\
  \texttt{lzh0212@g.ucla.edu} \\
  \And
  Haoyu Yang \\
  NVIDIA Research \\
  \texttt{haoyuy@nvidia.com} \\
  \And
  Mark Ho \\
  NVIDIA Research \\
  \texttt{chiatungh@nvidia.com} \\
  \And
  Yi-Chen Lu \\
  NVIDIA Research \\
  \texttt{yilu@nvidia.com} \\
  \And
  Wen-Hao Liu \\
  NVIDIA Research \\
  \texttt{wenhliu@nvidia.com} \\
  \And
  Shou-De Lin \\
  National Taiwan University \\
  \texttt{sdlin@csie.ntu.edu.tw} \\
  \And
  Mark Ren \\
  NVIDIA Research \\
  \texttt{haoxingr@nvidia.com} \\
}

\begin{document}

\maketitle

\begin{abstract}
Modern chip physical design relies heavily on Electronic Design Automation (EDA) tools, which often struggle to provide interpretable feedback or actionable guidance for improving routing congestion. In this work, we introduce a Multimodal Large Language Model Assistant (MLLMA) that bridges this gap by not only predicting congestion but also delivering human-interpretable design suggestions. Our method combines automated feature generation through MLLM-guided genetic prompting with an interpretable preference learning framework that models congestion-relevant tradeoffs across visual, tabular, and textual inputs. We compile these insights into a "Design Suggestion Deck" that surfaces the most influential layout features and proposes targeted optimizations. Experiments on the CircuitNet benchmark demonstrate that our approach outperforms existing models on both accuracy and explainability. Additionally, our design suggestion guidance case study and qualitative analyses confirm that the learned preferences align with real-world design principles and are actionable for engineers. This work highlights the potential of MLLMs as interactive assistants for interpretable and context-aware physical design optimization.
\end{abstract}

\section{Introduction}
\label{sec:intro}

% intro
The field of chip physical design encompasses critical prediction tasks such as congestion, timing, Design Rule Check (DRC), and IR Drop. These tasks, traditionally tackled using Electronic Design Automation (EDA) tools, often suffer from slow processing speeds and limited ability to provide actionable guidance for design improvement. Existing approaches using image-to-image translation models focus primarily on prediction accuracy but lack interpretability and the ability to suggest corrective measures. This paper explores the potential of Multimodal Large Language Models (MLLMs) in providing interpretable predictions along with actionable suggestions to assist engineers in optimizing their designs.
Recent advances in foundation models and instruction-tuned Large Language Models (LLMs) have led to impressive progress in tasks that require reasoning, code generation, and interactive assistance. However, their application to chip physical design remains largely underexplored. In particular, current EDA solutions lack the capability to provide real-time, explainable, and multimodal assistance that integrates layout visuals, tabular metrics, and design constraints into a unified decision-making process. We argue that a Multimodal Language Model Assistant (MLLMA) tailored to chip physical design can fill this gap.

% motivation
The motivation for this research is to address the gap between prediction accuracy and interpretability in physical design tasks. While current models are adept at predicting various metrics such as congestion and DRC violations, they fail to explain the underlying causes or offer corrective measures. By leveraging multimodal LLMs, we aim to move beyond mere prediction and offer a framework that provides both predictions and accompanying explanations. These explanations can be compiled into a "Design Suggestion Deck" which assists engineers by presenting comprehensible, human-readable reasoning to guide their design decisions.
% research problem
The primary research challenge is developing an MLLM-based agent framework capable of processing multimodal data inputs, including geometric images, tabular features, and circuit graphs, and providing interpretable predictions with actionable design suggestions. Furthermore, we aim to compare this approach against baseline models to demonstrate its effectiveness in improving design quality and interpretability.

% method
Our method consists of two stages: (1) Automate Feature Engineering and Generation, and (2) Interpretable Preferences Learning to Design Suggestion Deck.
In the first stage, a Multimodal Large Language Model (MLLM) agent generates and extracts numeric features from Macro Region, RUDY, and RUDY pin images and text-based features from configuration and logs. Inspired by the Genetic Instruct~\citep{majumdar2024genetic}, a genetic algorithm incorporating mutation and crossover is applied to expand the multimodal feature pool, with a deduplication process ensuring uniqueness. Feature importance and cross validation scores are evaluated using a Random Forest model to guide feature selection and refinement.
The second stage uses these features and their importance scores to create a Design Suggestion Deck. The MLLM agent translates important features into interpretable rules, providing actionable guidance for chip design optimization.

% experiment

% contribution
Our contributions are summarized as follows:

\begin{itemize}
    \item Development of a multimodal LLM framework capable of handling geometric image features, tabular data, and circuit graph inputs for prediction tasks in physical design.
    \item Design of an interpretable preference learning approach that generates interpretable predictions and actionable suggestions compiled into a "Design Suggestion Deck".
    \item Evaluation of the proposed framework against the baseline models that demonstrate strong improvements in design quality and efficiency.
\end{itemize}

% The remainder of this paper is structured as follows. In Section

\section{Related Work}
\label{sec:related}

\subsection{Multimodal Language Model Assistant}
\label{sec:related:mllm}

Recent works such as Mipha~\citep{zhu2024mipha}, MM-ReAct~\citep{yang2023mm}, and MultiModal-GPT~\citep{gong2023multimodal} demonstrate that aligning visual inputs with language reasoning modules allows models to solve complex real-world tasks. Mipha, for example, focuses on efficiency by enabling multi-turn multimodal dialogue with domain grounding, while MM-ReAct leverages reasoning traces and visual memory for tool-augmented perception.

In the context of chip design, prior efforts such as Chip-Chat~\citep{blocklove2023chip} investigate interactive co-design with conversational LLMs for hardware description generation, while ChipNeMo~\citep{liu2023chipnemo} introduces domain-adapted LLMs for industrial chip tasks including EDA script generation and bug analysis. Unlike these works, which primarily focus on language-to-code translation or domain adaptation, our approach leverages MLLMs for spatially grounded reasoning over fine-grained layout features such as routing demand and macro placement. Additionally, we explicitly model design tradeoffs through learned interpretable preferences, extending the role of LLMs from passive code generators to interactive design assistants capable of justification and feedback.

\subsection{Congestion Map Prediction}
\label{sec:related:congestion}

ClusterNet~\citep{min2023clusternet} focuses on predicting routing congestion caused by netlist topology using netlist clustering and Graph Neural Networks (GNNs). By employing the Leiden algorithm to generate cohesive clusters and developing a GNN-based model to generate cluster embeddings, ClusterNet achieves improved prediction performance and congestion optimization without requiring placement and routing (P\&R) tools.
CircuitNet~\citep{chai2023circuitnet,jiang2023circuitnet} provides an open-source dataset for machine learning applications in VLSI CAD, facilitating the development and benchmarking of ML models for tasks like congestion prediction and design rule check (DRC) violation prediction. The dataset includes diverse samples collected from various chip designs and supports comprehensive data, including routability, timing, and power metrics.
CircuitFormer~\citep{zou2023circuit} introduces a novel approach by treating circuit components as point clouds and utilizing Transformer-based point cloud perception methods for feature extraction. This method enables direct feature extraction from raw data without any preprocessing, allows for end-to-end training, and achieves state-of-the-art performance in congestion prediction tasks on both the CircuitNet and ISPD2015 datasets, as well as in DRC violation prediction tasks on the CircuitNet dataset.
MPGD~\citep{yang2024optimizing} proposes Mini-Pixel Batch Gradient Descent, a plug-and-play optimization algorithm that focuses on the most informative entries in the congestion map. By integrating this method into predictive AI models for physical design flows, MPGD enhances congestion prediction accuracy and efficiency, demonstrating its effectiveness in optimizing chip design processes.

\section{Methodology}
\label{sec:method}

Our proposed MLLM assistant consists of two core components:  
(1) automated feature generation and engineering, and  
(2) interpretable preference modeling for generating actionable design suggestions.

\subsection{Automated Feature Generation and Engineering}
\label{sec:method:feature}

\begin{figure}[t]
\centering
\makebox[\textwidth][c]{%
    \includegraphics[width=1.1\linewidth]{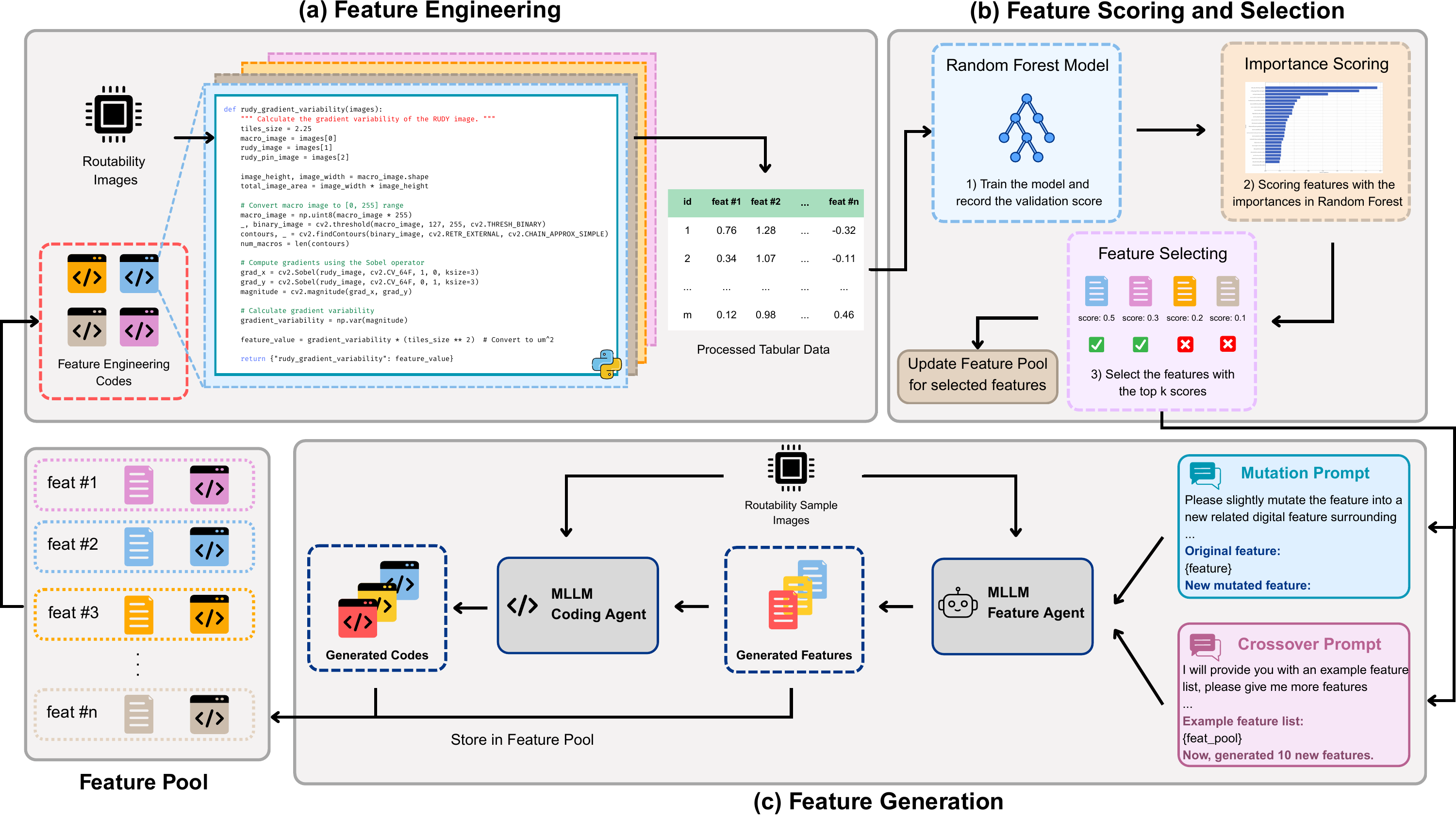}
}
\caption{Overview of the Genetic Instruct framework. (a) Spatial metrics are extracted from routability images using domain-specific rules. (b) A Random Forest ranks and selects key features. (c) MLLM agents iteratively generate and mutate features via code, enriching the feature pool to improve downstream performance.}
\label{fig:genetic-instruct}
\end{figure}

Figure~\ref{fig:genetic-instruct} illustrates our pipeline for automated feature generation and engineering. The approach integrates domain-driven initial features, MLLM-guided mutation, and a feedback loop to iteratively expand and refine a pool of predictive and interpretable features.

\subsubsection{Initial Feature Types}

We begin with two categories of raw features for routing congestion prediction:

\textbf{Image-based layout features:}  
\textit{Macro Region} encodes macro and cell placement for global layout context.  
\textit{RUDY (Routing Demand)} highlights routing density hotspots.  
\textit{RUDY Pin Images} capture local pin density patterns affecting routing complexity.

\textbf{Text-based configuration features:}  
Logs from EDA tools include placement constraints, blockages, net priorities, and timing directives. These symbolic features complement spatial representations with interpretable design signals.

% \begin{itemize}
%   \item \textbf{Image-Based Layout Features:}
%   \begin{itemize}
%     \item \textbf{Macro Region:} Captures the overall spatial layout of macros and standard cells, providing global context for placement-related congestion.
%     \item \textbf{RUDY (Routing Demand):} Highlights regions with high routing utilization, helping identify potential hotspots in the design.
%     \item \textbf{RUDY Pin Images:} Encodes pin density distributions, offering localized insight into areas prone to routing complexity.
%   \end{itemize}

%   \item \textbf{Text-Based Configuration Features:}
%   \begin{itemize}
%     \item \textbf{Routing Logs and Configuration:} Extracted from EDA tool output during global routing, these include placement constraints, blockages, net priorities, and timing-aware routing intents. They offer symbolic context that complements the spatial layout and helps guide optimization.
%   \end{itemize}
% \end{itemize}

\subsubsection{Genetic Instruct Feature Generation}

We employ a Genetic-Instruct framework~\citep{majumdar2024genetic} to iteratively evolve and expand feature sets using MLLMs. The process consists of initialization, evaluation, variation, and deduplication steps:

\textbf{Initialization:}  
The feature pool is seeded with handcrafted features that map visually and semantically to congestion causes. Each feature includes a human-readable name and description. MLLMs are prompted to generate extraction code, producing tabular data for training. Congestion severity is labeled using the average congestion value in the 20 most-congested grid cells of each sample.

\textbf{Evaluation and Selection:}  
A Random Forest model is trained on the extracted features and labels. The model’s feature importance scores act as the fitness function, guiding the pruning of low-impact features. We retain the top $k{=}20$ candidates for the next generation.

\textbf{Crossover and Mutation:}  
New features are created by prompting MLLMs with few-shot examples drawn from high-ranking features (crossover), or by mutating individual features with probabilities based on a rank-sensitive version of the Ebbinghaus forgetting curve. Specifically, mutation probability is defined as: $M = e^{-\frac{r}{N}}$, where \( r \) is the feature's rank and \( N \) is the total number of features. Higher-ranked features have greater mutation probability, encouraging diversity while maintaining quality.

\textbf{Deduplication:}  
To avoid redundancy, newly proposed features are compared against the existing pool using a Deduplicator-MLLM. It evaluates semantic similarity in descriptions and removes duplicates with reasoned justifications.

This process iteratively refines the feature space, producing a diverse, interpretable, and model-relevant set of features for downstream congestion prediction. Some generated feature example is shown in Figure~\ref{fig:feature-pool-example} and the complete list of generated feature pool is in Appendix~\ref{app:feature-pool}.

\begin{figure}[h]
\begin{tcolorbox}[title=Feature Pool Examples]
\footnotesize
\{\texttt{macro\_compactness\_index}: "a measure of how closely packed the macros are, potentially affecting routing paths and congestion", \\
\texttt{clustered\_macro\_distance\_std}: "the standard deviation of distances between clustered groups of macros", \\
\texttt{rudy\_pin\_clustering\_coefficient}: "a measure of how many rudy pins cluster together relative to the total number of rudy pins", \\
\texttt{macro\_density\_gradient}: "the change in macro density across the layout, impacting local congestion", ...\}
\end{tcolorbox}
% \texttt{macro\_aspect\_ratio\_variance}: "the variance in aspect ratios of macros, indicating potential alignment and spacing issues that may impact congestion", ...\}
\caption{Feature Pool Examples}
\label{fig:feature-pool-example}
\vspace{-0.5cm}
\end{figure}

\subsubsection{Automated Feature Engineering}

We employ Multimodal Large Language Models (MLLMs) to automate feature transformation by generating code snippets tailored to the characteristics of input features and target objectives. The generated code is then executed in a controlled environment, where inputs, outputs, and errors are systematically handled to ensure robustness. This automation streamlines the feature engineering process, improving both efficiency and scalability in modeling routing congestion.

\subsection{Interpretable Preferences to Design Suggestions}
\label{sec:method:interpretable}

\begin{figure}[t]
\centering
\makebox[\textwidth][c]{%
\includegraphics[width=1\linewidth]{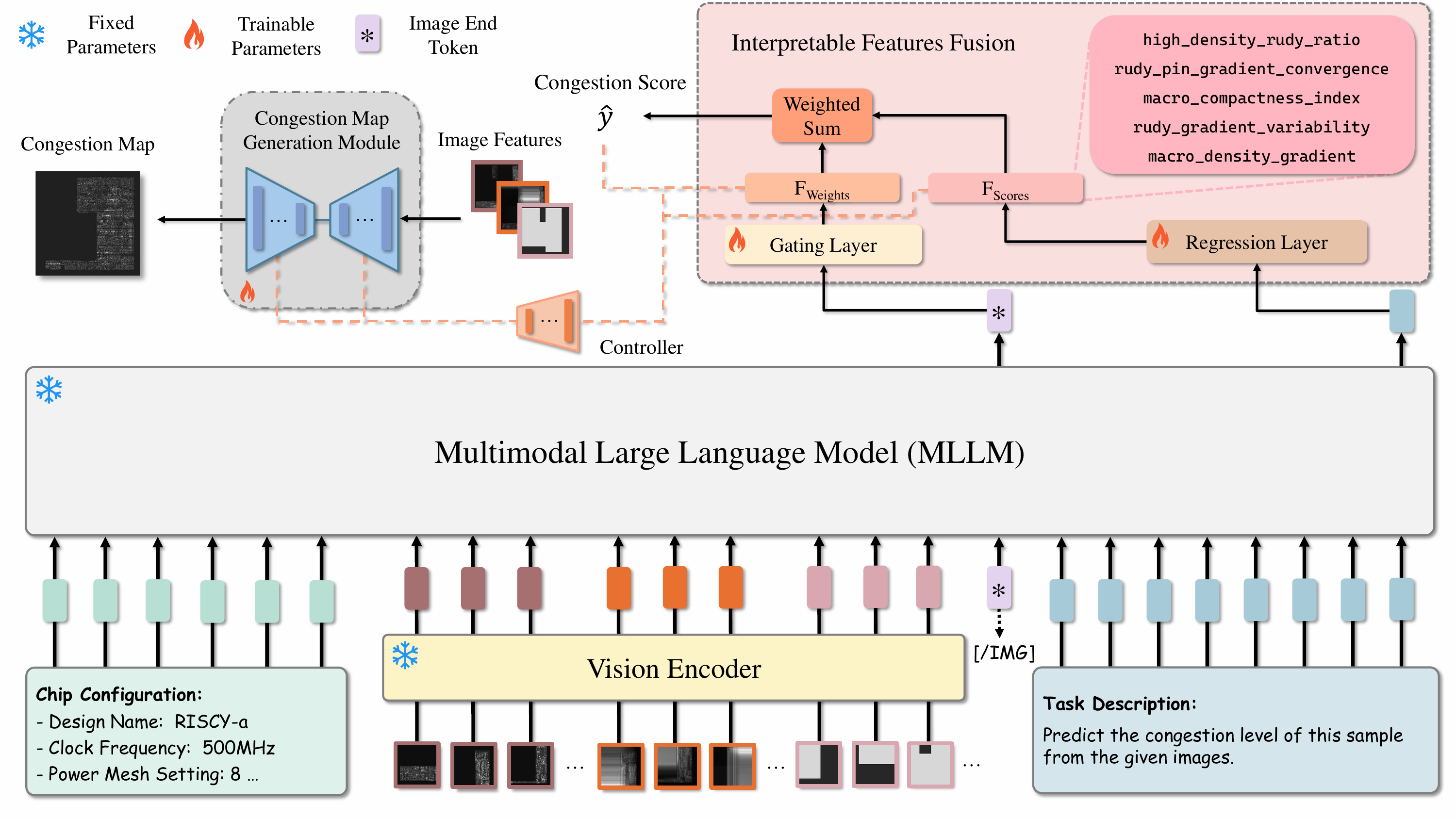}
}
\caption{
Overview of our interpretable preference modeling framework. The model leverages interpretable features identified in Phase 1 as preference signals to guide congestion prediction and design suggestion generation. A multimodal architecture integrates layout images and task prompts, predicts feature values, learns objective weights via a gating mechanism, and generates both scalar congestion scores and spatial congestion maps. This enables transparent and actionable feedback for physical design refinement.
}
\end{figure}

To bridge the gap between predictive modeling and actionable design insights, we develop a methodology that directly uses the features discovered in Phase 1 as interpretable preference indicators, enabling their translation into concrete and actionable design suggestions.

\subsubsection{Interpretable Preferences and Model Architecture}

Inspired by the ArmoRM framework \citep{wang2024interpretable}, we formulate design evaluation as a multi-objective task, addressing both scalar congestion prediction and spatial congestion map generation. Central to our approach is the reuse of the feature pool generated in Phase 1 via genetic instruction prompting. These features serve as interpretable preference indicators—each representing a physically meaningful and actionable design attribute such as pin clustering, macro placement density, or region-based congestion asymmetry.

Our model adopts a multimodal LLM backbone based on MiniCPM~\citep{yao2024minicpm}. Design configuration images are tokenized as vision inputs, while the task descriptions are encoded as text tokens. We then apply a regression layer trained on the last tokens to predict interpretable feature values, and a gating layer trained on vision tokens to learn objective weights based on layout context. The gating layer dynamically assigns importance to each interpretable feature dimension, effectively scalarizing the multi-objective signal into a preference-aware score.
This architecture enables both granular score decomposition and end-to-end learning across objectives. The model not only predicts congestion values but also generates pixel-level congestion maps, providing actionable, interpretable guidance to support iterative physical design refinement. 

In the congestion map generation stage, a conditional U-Net architecture is employed for the image-to-image translation task, where the input consists of image-based layout features and the output is the predicted congestion map. To incorporate the interpretable features as conditional information, we explore two representation methods: numerical vectors and text-based descriptions. Through the ablation studies in Appendix~\ref{app:ablation}, we find that using interpretable features with gating weights in a text-based representation achieves better performance, demonstrating their effectiveness in providing semantic information to guide the generation of congestion maps.

\subsubsection{Design Suggestions}

Using the interpretable preference signals, we generate targeted design suggestions focused on improving the most critical objectives. Since the model reasons over features that are both informative and physically grounded (e.g., pin density imbalance, macro boundary interactions), its suggestions are inherently actionable. For example, in cases where congestion is identified as a dominant issue, the model might recommend reducing cluster density in specific regions or modifying routing constraints. Each suggestion is directly traceable to feature-level insights, ensuring that design decisions are not only effective but also well-justified.

\section{Experiment}
\label{sec:exp}

Details of the experimental setup and hyperparameters are provided in Appendix~\ref{app:setup}.

\subsection{Dataset}
\label{sec:exp:dataset}
The CircuitNet dataset~\citep{jiang2023circuitnet,chai2023circuitnet} serves as the foundation for this study. It is a large-scale, open-source benchmark comprising over 10,000 synthesized chip designs—including CPUs, GPUs, and AI accelerators—fabricated using the 14nm FinFET process. It provides two types of features: image-like layout representations and text-based features that extracted from configuration log. Details on preprocessing and data augmentation are included in Appendix~\ref{app:setup}.

\subsection{Baselines}
\label{sec:exp:baselines}

% We compare our method with the following state-of-the-art congestion prediction models:
% \begin{itemize}
%     \item \textbf{GPDL}~\citep{chai2023circuitnet}: A CNN-based model trained on layout images to predict congestion maps. GPDL uses spatial convolutions to capture local routing patterns but lacks interpretable preference modeling.
%     \item \textbf{CircuitFormer}~\citep{zou2023circuit}: A Transformer-based point cloud model that learns spatial features from raw layout coordinates. CircuitFormer achieves strong performance on various congestion prediction benchmarks and serves as a competitive baseline for layout understanding.
% \end{itemize}
We compare our method against three state-of-the-art congestion prediction models. \textbf{GPDL}~\citep{chai2023circuitnet} is a CNN-based architecture that predicts congestion maps from layout images using spatial convolutions to model local routing patterns. \textbf{CircuitFormer}~\citep{zou2023circuit} is a Transformer-based model operating on point cloud data, learning spatial relationships from raw layout coordinates. \textbf{MPGD}~\citep{yang2024optimizing} introduces Mini-Pixel Batch Gradient Descent that targets the most informative regions of the congestion map.

\subsection{Evaluation Metrics}
\label{sec:exp:eval}
Each model is trained to output a pixel-level prediction of congestion, which is then compared against the ground-truth congestion heatmap. We adopt standard metrics used in prior work to evaluate congestion prediction quality. These include SSIM for perceptual similarity, NRMSE and PeakNRMSE for error measurement, and PLCC, SRCC, and KRCC for correlation analysis. A detailed explanation and formal definitions of each metric are provided in Appendix~\ref{app:eval-metrics}.

\subsection{Experiment Results}
\label{sec:exp:results}

\begin{figure}[ht]
    \centering
    \makebox[\textwidth]{%
    \resizebox{1\linewidth}{!}{%
        \begin{tabular}{c}
            \includegraphics[trim=15 40 15 28, clip]{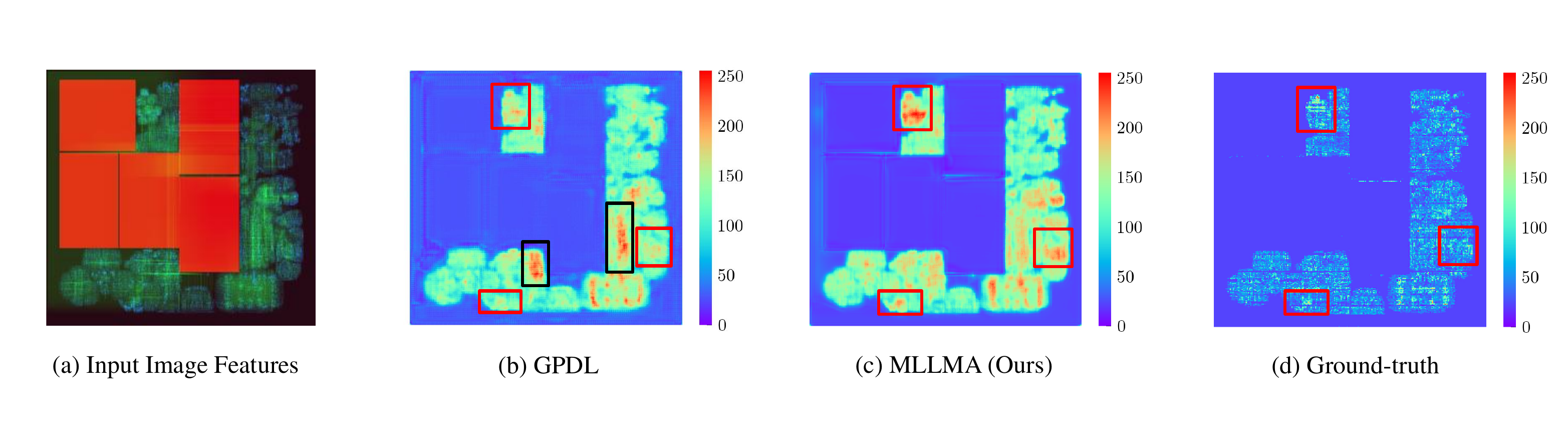}
        \end{tabular}
    }
    }
    \caption{
    Qualitative comparison of congestion prediction heatmaps between the GPDL baseline and our proposed multimodal model, MLLMA. Red bounding boxes highlight congestion hotspots where MLLMA more accurately captures the spatial patterns and severity of congestion compared to the ground truth. Black bounding boxes indicate regions where GPDL fails to localize congestion accurately.
    }
    \vspace{-1cm}
    \label{fig:qualitative-results}
\end{figure}

Table~\ref{tab:main-results} shows that our model consistently outperforms baselines in SSIM, NRMSE, and peak error metrics, validating the benefit of interpretable preference modeling and multi-objective supervision.
Our qualitative results in Figure~\ref{fig:qualitative-results} further show that the predictions from our model more accurately capture fine-grained congestion boundaries compared to the baselines.

% \begin{table}[ht]
% \centering
% \caption{Congestion prediction performance comparison on CircuitNet.}
% \label{tab:combined}
% \begin{subtable}[t]{\textwidth}
% \centering
% \caption{SSIM, NRMSE, and peak NRMSE.}
% \label{tab:main-results-1}
% % \resizebox{\textwidth}{!}{%
% \begin{tabular}{c|c|c|ccccc}
% \toprule
% \multirow{2}{*}{\textbf{Method}} & \multirow{2}{*}{\textbf{SSIM ↑}} & \multirow{2}{*}{\textbf{NRMSE ↓}} & \multicolumn{5}{c}{\textbf{peak NRMSE ↓}} \\
%  & & & 0.5\% & 1\% & 2\% & 5\% & avg \\
% \midrule
% GPDL~\citep{chai2023circuitnet} & 0.773 & 0.047 & 0.441 & 0.323 & 0.236 & 0.155 & 0.289\\ 
% MPGD~\citep{yang2024optimizing} & 0.787 & 0.046 & 0.271 & 0.217 & 0.171 & 0.121 & 0.195\\
% \textbf{Ours} & \textbf{0.791} & \textbf{0.047} & \textbf{0.440} & \textbf{0.323} & 0.237 & 0.157 & \textbf{0.289}\\
% \bottomrule
% \end{tabular}%
% % }
% \end{subtable}
% \hfill
% \begin{subtable}[t]{\textwidth}
% \centering
% \caption{Correlation metrics.}
% \label{tab:main-results-2}
% % \resizebox{\textwidth}{!}{%
% \begin{tabular}{c|c|c|c}
% \toprule
% \textbf{Method} & \textbf{PLCC ↑} & \textbf{SRCC ↑} & \textbf{KRCC ↑} \\
% \midrule
% GPDL & 0.5032 & 0.5143 & 0.3787\\
% CircuitGNN & 0.3287 & 0.4483 & 0.3688 \\
% CircuitFormer~\citep{zou2023circuit} & 0.6374 & 0.5282 & 0.3935 \\
% \textbf{Ours} & \textbf{0.xxx} & \textbf{0.xxx} & \textbf{0.xxx} \\
% \bottomrule
% \end{tabular}%
% % }
% \end{subtable}
% \end{table}

\begin{table}[ht]
\centering
\caption{Congestion prediction performance on CircuitNet across pixel-based and correlation-based metrics.}
\label{tab:main-results}
\makebox[\textwidth]{ % ensures centering
\resizebox{1.1\textwidth}{!}{
\begin{tabular}{l|cc|ccccc|ccc}
\toprule
\multirow{2}{*}{\textbf{Method}} & \multirow{2}{*}{\textbf{SSIM ↑}} & \multirow{2}{*}{\textbf{NRMSE ↓}} & \multicolumn{5}{c|}{\textbf{Peak NRMSE ↓}} & \multicolumn{3}{c}{\textbf{Correlation Metrics}} \\
 & & & 0.5\% & 1\% & 2\% & 5\% & avg & PLCC ↑ & SRCC ↑ & KRCC ↑ \\
\midrule
GPDL~\citep{chai2023circuitnet}      & 0.773 & 0.047 & 0.441 & 0.323 & 0.236 & 0.155 & 0.289 & 0.5032 & 0.5143 & 0.3787 \\
MPGD~\citep{yang2024optimizing}      & 0.787 & 0.046 & 0.271 & 0.217 & 0.171 & 0.121 & 0.195 & --     & --     & --     \\
CircuitGNN                           & --    & --    & --    & --    & --    & --    & --    & 0.3287 & 0.4483 & 0.3688 \\
CircuitFormer~\citep{zou2023circuit} & --    & --    & --    & --    & --    & --    & --    & 0.6374 & \textbf{0.5282} & 0.3935 \\
\textbf{Ours}                        & \textbf{0.807} & \textbf{0.045} & \textbf{0.263} & \textbf{0.205} & \textbf{0.169} & \textbf{0.118} & \textbf{0.189} & \textbf{0.6452} & 0.5233 & \textbf{0.3941} \\
\bottomrule
\end{tabular}
}}
\vspace{-0.3cm}
\end{table}

\subsection{Ablation Study}
\label{sec:ablation}

% phase 1
% GDPL
% naive prompting
% Genetic instruct

% phase 2
% feature ablation: image/config
% preference feature ablation: numeric/text

To understand which components contribute most to our model's effectiveness, we conduct a two-phase ablation study.

\paragraph{Phase 1: Ablation Study on Feature Generation Pipeline}  
This phase evaluates the necessity and effectiveness of our proposed \textbf{Genetic Instruct prompting strategy} for MLLMs. We compare it against two variants: a crossover-only version that recombines prompt elements and a mutation-only version that introduces random prompt variations. Our Genetic Instruct method integrates both crossover and mutation with evolutionary selection, and achieves significantly better feature quality for downstream congestion prediction.
As shown in Table~\ref{tab:ablation-prompt}, the Genetic Instruct strategy leads to substantial performance gains, demonstrating that careful prompt optimization is crucial when leveraging MLLMs for physical design tasks.

\paragraph{Phase 2: Ablation Study on Interpretable Preference Pipeline}  
This phase investigates which components of our interpretable preference modeling pipeline are most critical to performance. We ablate four key aspects: the input modalities, the loss functions, the congestion map generation modules, and the interpretable preference features used for gating mechanism. Each removal leads to varying degrees of performance degradation, highlighting the importance of both structural and semantic components in learning meaningful preferences.
A subset of the ablation results is summarized in Table~\ref{tab:ablation-results-subset}, with additional details provided in Appendix~\ref{app:ablation}. These findings confirm that the learned preference signals are not only interpretable but also integral to model performance. Notably, removing structured inputs or disabling the gating mechanism leads to substantial degradation, underscoring the critical role of each component in modeling contextual and actionable design preferences.

\begin{table}[t]
\centering
\caption{Ablation Studies.}
\label{tab:ablation-results-subset}
\makebox[\textwidth]{%
\resizebox{1.15\textwidth}{!}{%
\begin{tabular}[t]{@{}c@{}c@{}}
%%%%%%%%%%%%%%%%%%%%%%%%%%%%%%%%%%%%%%%%%%%%%%%%%%%%%%%%
\begin{tabular}[t]{@{}l@{}}
(a) Ablation on feature generation pipeline. \\
\begin{tabular}{l|ccc|ccc}
\toprule
\multirow{2}{*}{\textbf{Pipeline Variant}} & \multicolumn{3}{c|}{\textbf{zero-riscy-a}} & \multicolumn{3}{c}{\textbf{zero-riscy-b}} \\
& \textbf{PLCC ↑} & \textbf{SRCC ↑} & \textbf{KRCC ↑} & \textbf{PLCC ↑} & \textbf{SRCC ↑} & \textbf{KRCC ↑} \\
\midrule
Crossover-Only                    & 0.630 & 0.674 & 0.492  & 0.589 &\textbf{ 0.686} & \textbf{0.493}\\
Mutation-Only              & 0.353 & 0.485 & 0.348 & 0.469 & 0.462 & 0.314 \\
Genetic Instruct        & \textbf{0.705} & \textbf{0.786} & \textbf{0.595} & \textbf{0.599} & 0.591 & 0.412 \\
\bottomrule
\end{tabular}
\end{tabular}
&
%%%%%%%%%%%%%%%%%%%%%%%%%%%%%%%%%%%%%%%%%%%%%%%%%%%%%%%%
\renewcommand{\arraystretch}{1.23}
\begin{tabular}[t]{@{}l@{}}
(b) Ablation on interpretable features. \\
\begin{tabular}{l|c|c}
\toprule
\textbf{Feature Source Modality} & \textbf{SSIM ↑} & \textbf{NRMSE ↓} \\
\midrule
Image-Based & \textbf{0.791} & 0.047 \\
Log-Based   & 0.764 & 0.052 \\
Mixed (Image + Log)      & 0.789 & \textbf{0.046} \\
\bottomrule
\end{tabular}
\end{tabular}
%%%%%%%%%%%%%%%%%%%%%%%%%%%%%%%%%%%%%%%%%%%%%%%%%%%%%%%%
\end{tabular}
}}
\end{table}

\section{Analysis}
\label{sec:analysis}

\subsection{Feature Attribution Analysis}
\label{sec:analysis:attribution}

Each design feature is associated with a scalar gating value representing its relative importance or preference weight. To interpret the sign of these values:
These interpretations allow us to understand how each feature drives model predictions in different design contexts.
We observe that low-congestion samples tend to exhibit high values on the \texttt{macro\_rudy\_boundary\_interaction\_index} feature, as illustrated in Table~\ref{tab:feature-correlation}. This trend is consistent across multiple instances, with a negative Pearson correlation of $-0.72$ between this feature and the congestion score.

Similarly, features such as \texttt{rudy\_pin\_clustering\_coefficient} and \texttt{rudy\allowbreak\_pin\allowbreak\_compaction\allowbreak\_ratio} tend to have high values in high-congestion samples, showing a positive correlation. This suggests that pin-related density features are dominant contributors to congestion under high-load scenarios.

To further support interpretability, we conduct a detailed breakdown analysis covering the value distribution, gating value, and associated congestion level for every major design feature. This comprehensive analysis is included in Appendix~\ref{app:attribution-breakdown} and Figure~\ref{app:feature-attribution}, where we also include additional observations and insights derived from these patterns.

\begin{table}[ht]
    \centering
    \includegraphics[width=\linewidth]{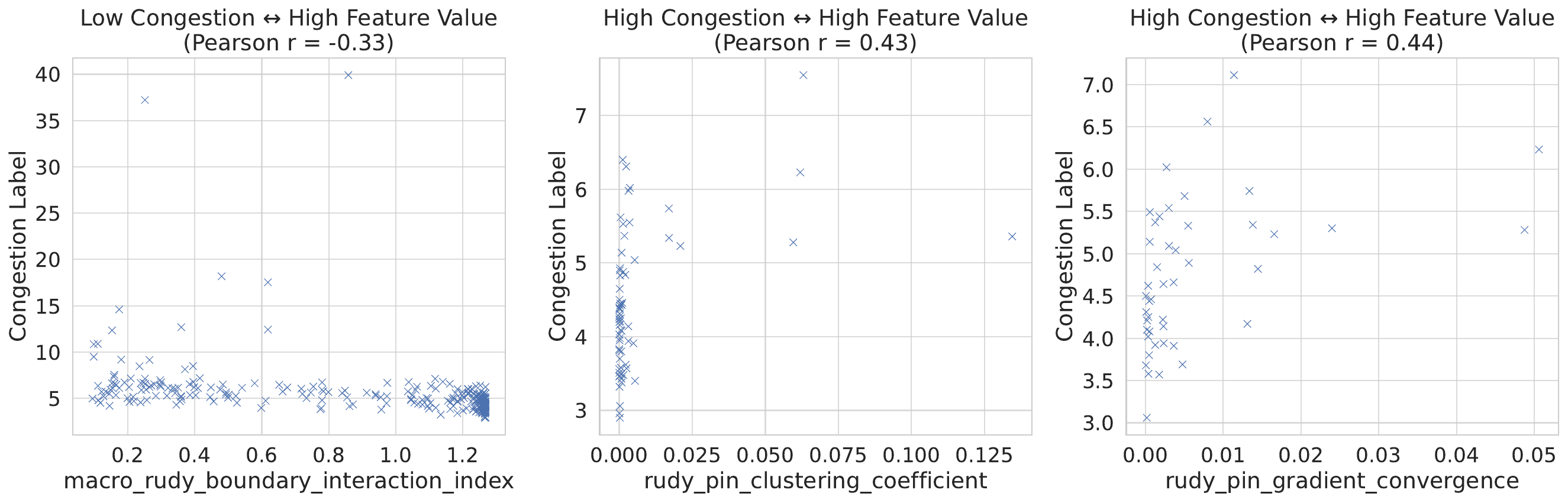}
    \caption{
    Scatter plots showing the relationship between key layout features and congestion labels.
    (Left) \texttt{macro\_rudy\_boundary\_interaction\_index} is negatively correlated with congestion ($r = -0.33$), indicating that low-congestion samples tend to exhibit higher values in this macro-level feature.
    (Middle) \texttt{rudy\_pin\_clustering\_coefficient} shows a moderate positive correlation with congestion ($r = 0.43$), suggesting that dense pin clusters contribute to congestion under high-load conditions.
    (Right) \texttt{rudy\_pin\_gradient\_convergence} similarly displays a positive trend ($r = 0.44$), further highlighting the influence of pin-level spatial patterns on congestion outcomes.
    }
    \label{tab:feature-correlation}
\end{table}

\subsection{Design Suggestion Guidance Leads To Lower Congestion}
\label{sec:analysis:case-study}

\begin{figure}[ht]
    \centering
    \begin{subfigure}[t]{0.48\linewidth}
        \centering
        \vspace{0pt}
        \includegraphics[width=\linewidth]{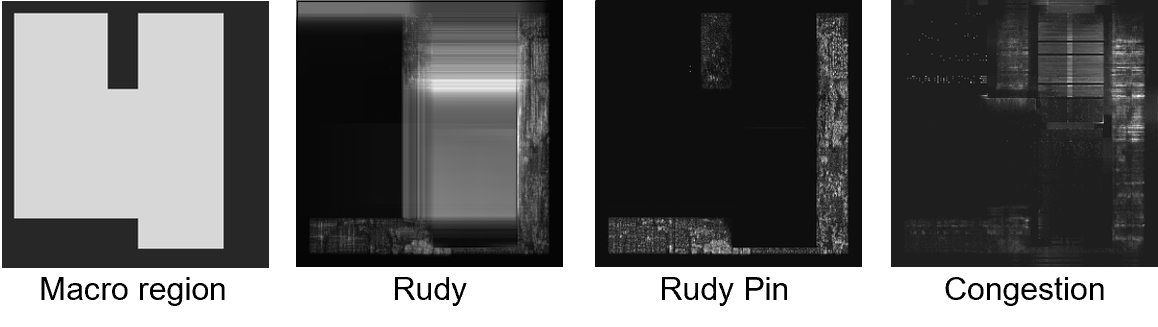}
        \caption{Before adjustment.}
        \label{fig:before}
    \end{subfigure}
    \hfill
    \begin{subfigure}[t]{0.48\linewidth}
        \centering
        \vspace{0pt}
        \includegraphics[width=\linewidth]{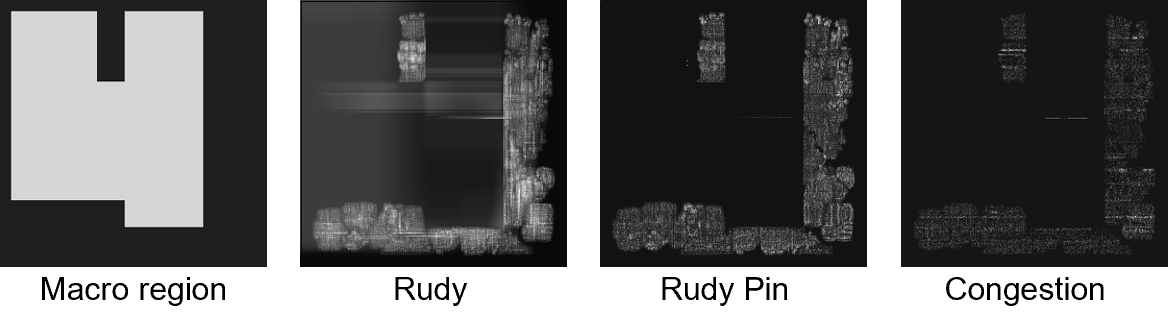}
        \caption{After adjustment.}
        \label{fig:after}
    \end{subfigure}
    \caption{By following the top model-attributed features and adjusting actionable design parameters, the resulting design exhibits significantly lower predicted and actual congestion.}
    \label{fig:case-study}
\end{figure}

To evaluate the practical utility of the model’s interpretable outputs, we conduct a case study that demonstrates how applying model-guided feature adjustments can lead to a substantial reduction in routing congestion. Specifically, we use the model’s top-attributed features to inform actionable design changes and compare outcomes before and after applying these suggestions.
Figure~\ref{fig:case-study} presents one representative example. The left side shows a high-congestion design sample, along with the top-5 features contributing most to the model’s prediction. Based on these attributions, we systematically adjust feature values—particularly those with strong positive influence on congestion—toward more favorable levels. The adjusted design, shown on the right, exhibits a notable improvement.

This case highlights the effectiveness of the model's interpretable reasoning. The ability to trace predictions back to key design features enables not just post hoc understanding, but also actionable guidance. By following these learned preferences, designers can reduce congestion and iteratively improve design quality in practice.

\vspace{0.5em}
\noindent\textbf{Feature Attribution and Congestion Change Summary:}

\begin{table}[ht]
\centering
\caption{Top features before/after adjustment and their impact on congestion.}
\label{tab:feature-change-impact}
\resizebox{\linewidth}{!}{%
\begin{tabular}{l|c|c|c|c}
\toprule
\textbf{Feature} & \textbf{Before Value} & \textbf{After Value} & \textbf{Attribution (Before)} & \textbf{Congestion Impact} \\
\midrule
rudy\_pin\_gradient\_convergence & 0.3351 & 0.1202 & 0.1551 → 0.0001 & High \\
rudy\_pin\_clustering\_coefficient & 0.5913 & 0.3606 & 0.1409 → 0.0002 & ↓ \\
rudy\_pin\_compaction\_ratio & 0.4273 & 0.3035 & 0.1216 → 0.0001 & ↓ \\
high\_density\_rudy\_pin\_ratio & 0.3138 & 0.4867 & 0.0638 → 0.0003 & Mixed \\
demarcated\_macro\_proximity\_index & 0.3765 & -0.0030 & 0.0539 → 0.0000 & ↓ \\
\midrule
macro\_rudy\_boundary\_interaction\_index & 0.0849 & 0.0721 & 0.0321 → 1.2655 & High post-adjustment role \\
\bottomrule
\end{tabular}
}
\end{table}

\noindent
This case illustrates how interpreting the gating values and using them to guide feature adjustment can lead to measurable congestion reduction, demonstrating the interpretability and actionability of our model’s learned preferences.

We compute the consistency rate across a large number of matched sample pairs. A high agreement between the sign of the gating value and the direction of the label difference indicates that the learned preferences are interpretable and aligned with the optimization objective. Additionally, we collaborate with experienced chip designers and researchers to manually review selected cases, further validating that the gating values reflect meaningful and actionable design insights.

\subsection{Design Suggestion Qualitative Examples}
\label{sec:analysis:qualitative}

% To qualitatively interpret model predictions, we compare two designs with different congestion levels: 8277 (40.21) and 8243 (20.04). As shown in Figure~\ref{fig:qualitative-examples}, Design 8277 relies on a broad set of moderately important, pin-related features (e.g., clustering, compaction), while 8243 is influenced almost entirely by macro-RUDY boundary interactions.
% This example demonstrates that even with similar feature values, the importance differs by layout context—supporting our goal of design-specific interpretability.
% Full qualitative breakdown and diagnostic UI available at\footnote{\url{https://chatgpt.com/share/68131ad8-92d4-800d-9e8c-df1ddd6283ba}}

\begin{figure*}[ht]
    \centering
    \begin{minipage}[t]{0.49\linewidth}
        \vspace{0pt}
        \includegraphics[width=\linewidth, trim={0 40 0 0}, clip]{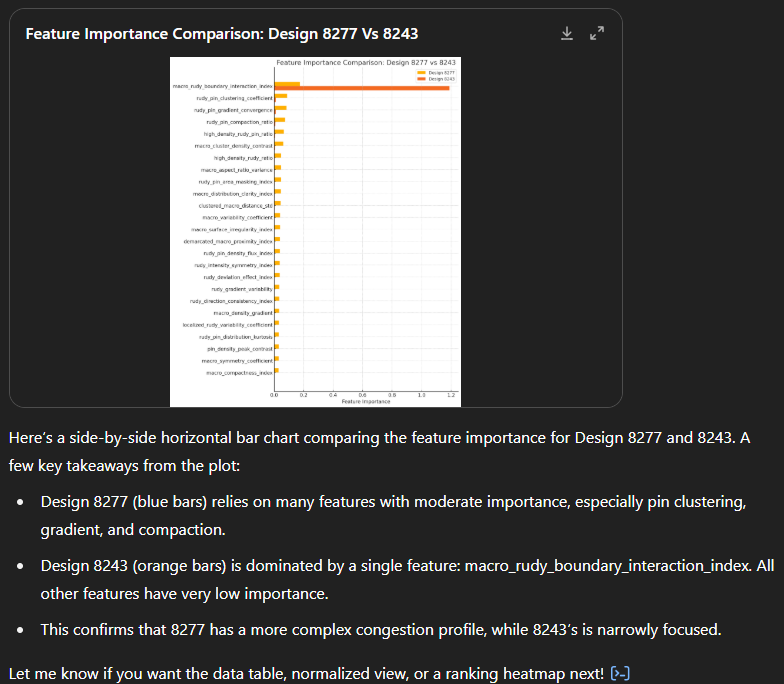}
    \end{minipage}
    \hfill
    \begin{minipage}[t]{0.49\linewidth}
        \vspace{0pt}
        \includegraphics[width=\linewidth]{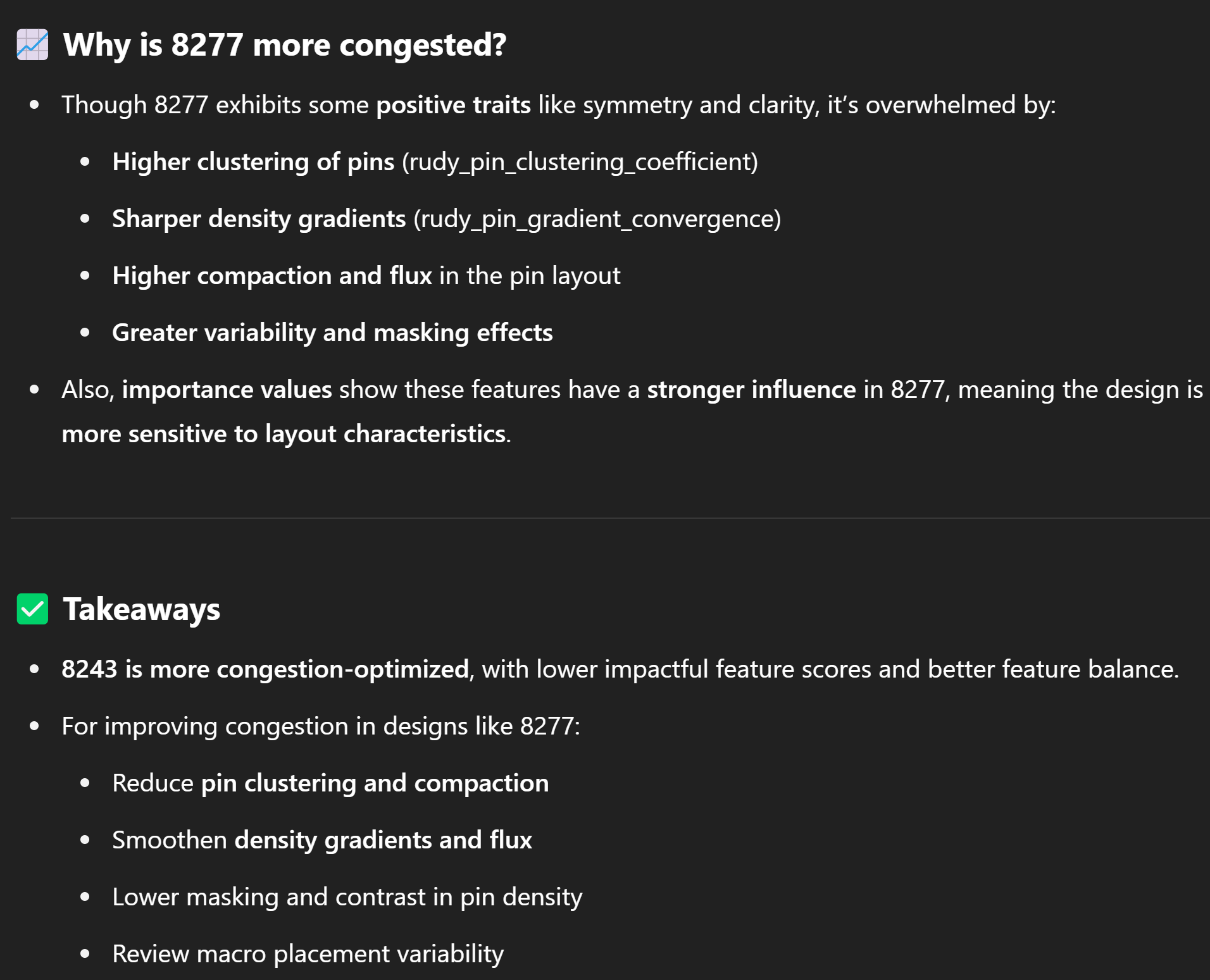}
    \end{minipage}
    \caption{Qualitative interpretation comparing Design 8277 and 8243. Left: Feature importance bar chart. Right: Root causes and actionable takeaways.}
    \label{fig:qualitative-examples}
\end{figure*}

We complement our quantitative and case-based evaluation with a set of qualitative examples that further illustrate how the model adapts its reasoning to context-specific designs. In Figure~\ref{fig:qualitative-examples}, we compare two designs—Design 8277 and Design 8243—exhibiting congestion scores of 40.21 and 20.04, respectively.
Despite having similar raw feature values in some cases, the relative importance assigned to each feature varies dramatically between the two layouts. Design 8277 shows elevated importance on multiple pin-related density features, suggesting that the layout suffers from excessive pin compaction and clustering. In contrast, Design 8243 is dominated by a single macro-level feature—\texttt{macro\_rudy\_boundary\_interaction\_index}—highlighting the positive influence of well-structured macro placement.
These examples demonstrate that our model not only identifies which features matter most, but also how their importance is modulated by spatial context. This supports our broader goal of providing interpretable, design-aware guidance that generalizes across different physical design scenarios.

% \begin{figure*}[ht]
%     \centering
%     \begin{minipage}[t]{0.32\linewidth}
%         \vspace{0pt}
%         \includegraphics[width=\linewidth]{Figs/qualitative_compare.png}
%     \end{minipage}
%     \hfill
%     \begin{minipage}[t]{0.32\linewidth}
%         \vspace{0pt}
%         \includegraphics[width=\linewidth, trim={0 90 0 0}, clip]{Figs/qualitative_8277_profile.png}
%     \end{minipage}
%     \hfill
%     \begin{minipage}[t]{0.32\linewidth}
%         \vspace{0pt}
%         \includegraphics[width=\linewidth]{Figs/qualitative_8243_profile.png}
%     \end{minipage}
%     \caption{Qualitative comparison of Design 8277 and 8243. Left: Feature importance comparison. Middle: 8277 profile—congestion influenced by pin-level density. Right: 8243 profile—dominated by macro-RUDY boundary effects.}
%     \label{fig:qualitative-examples}
% \end{figure*}

\section{Limitation}
\label{sec:limitation}

While our framework shows strong performance, several limitations remain. First, the current evaluation is limited to a RISC-V chip architectures and the CircuitNet dataset. Broader validation across different architectures and more diverse commercial designs is necessary for practical deployment. Second, our focus is confined to congestion prediction; extending the approach to other critical physical design objectives such as power, performance, area (PPA), IR drop, and timing closure remains an open direction.

\section{Conclusion}
\label{sec:conclusion}

We propose a multimodal LLM assistant for chip physical design that predicts congestion and provides interpretable guidance. By combining genetic prompting and preference-based explanation, our method delivers actionable design suggestions aligned with expert intuition. Experiments show state-of-the-art results on CircuitNet, and case studies confirm practical utility. Future work will expand to broader tasks and interactive, real-time, instruction-like design support.

\begin{ack}
\end{ack}

\bibliography{Styles/neurips_2025}
\clearpage
\appendix

\section{Automated Feature Generation and Feature Engineering Algorithm}

% Add Automated algorithm
\begin{algorithm}[ht]
\setstretch{1.15}
\caption{Automated Feature Generation and Feature Engineering}
  \label{alg::Automated Feature Generation and Feature Engineering}
  \begin{algorithmic}[1]
   \Require{$N$: Number of automated iterations \\
        $k$: Maximum size of feature pool \\ 
        $\mathcal{F}_{init}$: Initial set of hand-crafted features  \\
        $\mathcal{D}$: Raw images of training data \\
        $\mathcal{L}$: Congestion levels of training data \\
        $I_{s}$: Sample images for the inputs of MLLMs \\
        $P_{m}$: Probability of selecting mutation operation \\
        $P_{op}$: Probability distribution over the operations \{Mutation: $P_{m}$, Cross-Over: $1-P_{m}$\}
    }
   \Ensure{$\mathcal{F}_{total}$: Final feature pool,  $\mathcal{E}_{total}$: Final feature extraction codes} 
        
    \State Initialize $\mathcal{F}_{pool} \leftarrow \mathcal{F}_{init};$ 
    \State Initialize $\mathcal{E}_{code} \leftarrow CoderMLLM(I_s, \mathcal{F}_{pool});$
    
    \For {$r \leftarrow 1$ \textit{\textbf{to}} $N$}
        \State Initialize $T \leftarrow \{\mathcal{L}\};$
        \For {each feature extractor $\epsilon_{i}$ in $\mathcal{E}_{code}$}
            \State $T \leftarrow T \ \cup \ \epsilon_{i}(\mathcal{D});$
        \EndFor
         \State $R \leftarrow$ Train a Random Forest Regressor with tabular features $T$;
         \State $f \leftarrow$ Return the feature importance function in trained Random Forest Regressor $R$;
         \State $\mathcal{F}_{pool}, \mathcal{E}_{code} \leftarrow FeatureSelector(\mathcal{F}_{pool},\mathcal{E}_{code},f, \min(k, len(\mathcal{F}_{pool})));$
         \State $S_{op} \leftarrow$ Choose an operation from $P_{op}$;
         \If {$S_{op} = $ Mutation}
             \State $M \leftarrow RankSensitveFormula(\mathcal{F}_{pool},f);$ 
             \State $\mathcal{F}_{parents} \leftarrow$ Select parent features $\subset \mathcal{F}_{pool}$ from mutation probabilities $M$;
             \State $\mathcal{F}_{new} \leftarrow MutatorMLLM(I_s, \mathcal{F}_{parents});$
         \Else
             \State $\mathcal{F}_{new} \leftarrow CrossoverMLLM(I_s, \mathcal{F}_{pool});$
         \EndIf
         \State $\mathcal{F}_{new} \leftarrow DeduplicatorMLLM(I_s, \mathcal{F}_{pool}, \mathcal{F}_{new});$
         \State $\mathcal{E}_{new} \leftarrow CoderMLLM(I_s, \mathcal{F}_{new});$
         \State $\mathcal{F}_{pool} \leftarrow \mathcal{F}_{pool} \ \cup \ \mathcal{F}_{new}   ;$ 
         \State $\mathcal{E}_{code} \leftarrow \mathcal{E}_{code} \ \cup \ \mathcal{E}_{new}   ;$
    \EndFor

    \State $\mathcal{F}_{total}, \mathcal{E}_{total}  \leftarrow \mathcal{F}_{pool}, \mathcal{E}_{code};$  
   \end{algorithmic}
\end{algorithm}

\clearpage

\section{Evaluation Metric Definition}
\label{app:eval-metrics}

\begin{definition}[SSIM]
The \textbf{Structural Similarity Index Measure (SSIM)} evaluates the perceptual similarity between two images based on luminance, contrast, and structural information. It is defined as:
\[
\text{SSIM}(x, y) = \frac{(2\mu_x \mu_y + c_1)(\sigma_{xy} + c_2)}{(\mu_x^2 + \mu_y^2 + c_1)(\sigma_x^2 + \sigma_y^2 + c_2)},
\]
where $\mu$ denotes the mean, $\sigma^2$ the variance, and $\sigma_{xy}$ the covariance between prediction $x$ and ground truth $y$. Constants $c_1$ and $c_2$ stabilize the denominator.
\end{definition}

\vspace{2cm}

\begin{definition}[MSE, NRMSE, PeakNRMSE]
The \textbf{Mean Squared Error (MSE)} is defined as:
\[
\text{MSE}(x, y) = \frac{1}{n} \sum_{i=1}^{n} (x_i - y_i)^2.
\]
The \textbf{Normalized Root Mean Square Error (NRMSE)} scales the RMSE by the value range of the ground truth:
\[
\text{NRMSE}(x, y) = \frac{1}{\max(y) - \min(y)} \sqrt{\text{MSE}(x, y)}.
\]
The \textbf{PeakNRMSE} evaluates NRMSE restricted to the top-$k$ largest entries of both $x$ and $y$, emphasizing high-congestion regions that are most critical in physical design.
\end{definition}

\vspace{2cm}

\begin{definition}[Correlation-Based Metrics: PLCC, SRCC, KRCC]
These metrics evaluate statistical and rank-order correlation between prediction $x$ and ground truth $y$:
\begin{itemize}
    \item \textbf{PLCC (Pearson Linear Correlation Coefficient)} measures linear correlation:
    \[
    \text{PLCC}(x, y) = \frac{\text{Cov}(x, y)}{\sigma_x \sigma_y},
    \]
    where $\text{Cov}(x, y)$ is the covariance and $\sigma$ is the standard deviation.

    \item \textbf{SRCC (Spearman Rank-Order Correlation Coefficient)} computes PLCC on the rank-transformed values of $x$ and $y$, capturing monotonic relationships.

    \item \textbf{KRCC (Kendall Rank Correlation Coefficient)} compares the number of concordant and discordant pairs:
    \[
    \text{KRCC}(x, y) = \frac{n_c - n_d}{\binom{n}{2}},
    \]
    where $n_c$ and $n_d$ are the counts of concordant and discordant pairs among all $\binom{n}{2}$ possible pairs.
\end{itemize}
\end{definition}

\clearpage

\section{Feature Attribution Breakdown Across All Features}
\label{app:attribution-breakdown}

\begin{figure}[ht]
    \centering
    \begin{subfigure}[t]{0.75\textwidth}
        \centering
        \includegraphics[width=\linewidth]{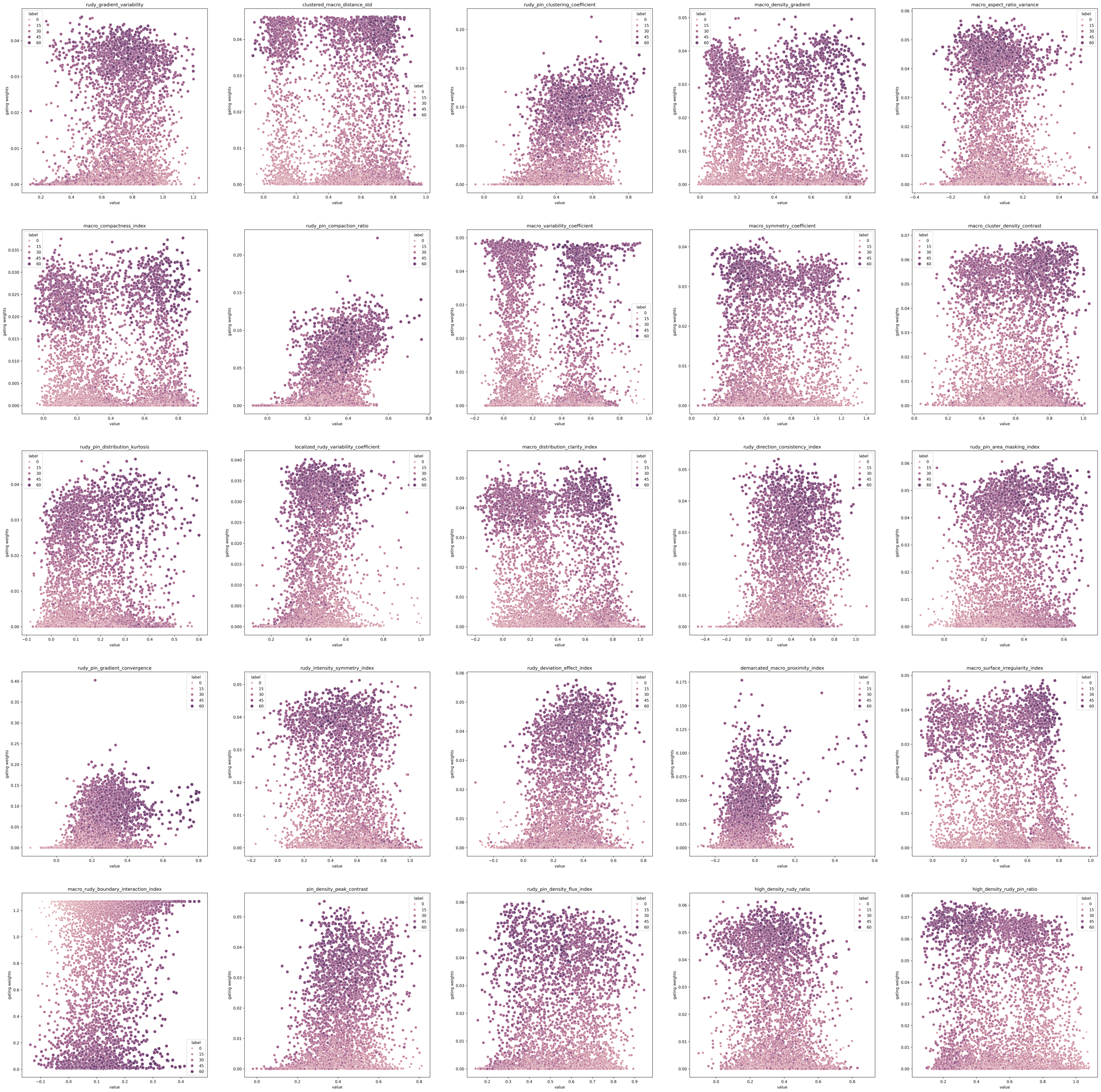}
        \caption{Gating values vs. feature values, colored by predicted congestion.}
        \label{fig:attribution-breakdown-pred}
    \end{subfigure}
    \hfill
    \begin{subfigure}[t]{0.75\textwidth}
        \centering
        \includegraphics[width=\linewidth]{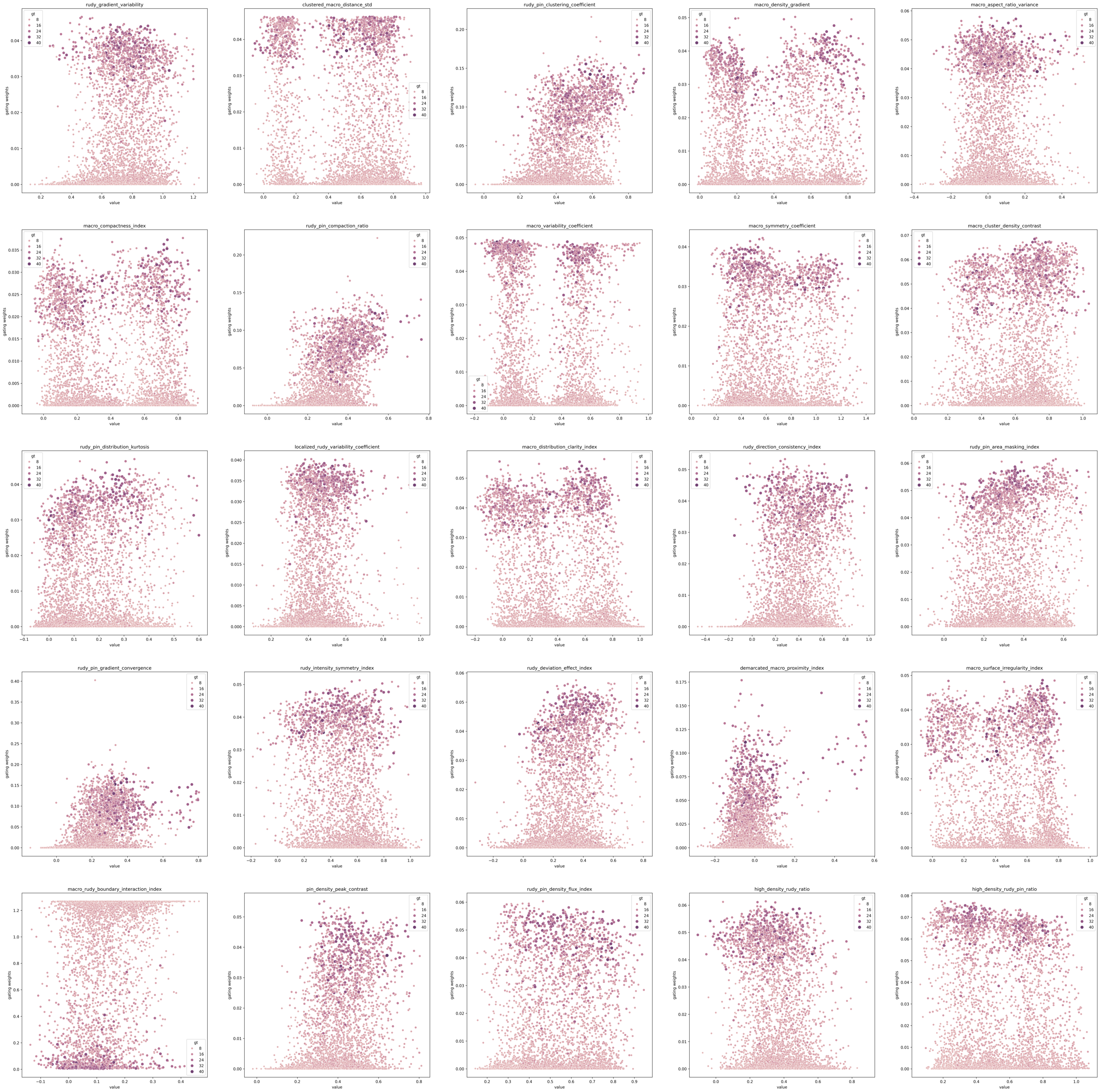}
        \caption{Gating values vs. feature values, colored by ground-truth congestion.}
        \label{fig:attribution-breakdown-label}
    \end{subfigure}
    \caption{Feature attribution breakdown across all 25 features. Each point represents a sample, colored by congestion score.}
    \label{fig:attribution-breakdown}
\end{figure}

To further examine the interpretability of the learned gating values, we analyze all 25 features used by the model. Figure~\ref{fig:attribution-breakdown} presents a comprehensive visualization, where each subplot corresponds to one feature. 

For each subplot:
\begin{itemize}
    \item The x-axis denotes the raw feature value.
    \item The y-axis shows the corresponding gating weight assigned by the model.
    \item Each point is colored by its associated congestion score, with darker colors indicating higher congestion.
    \item The left panel shows results from model \textit{predictions}, and the right panel shows alignment with \textit{ground-truth labels}.
\end{itemize}

\paragraph{Observation 1: Prediction-Label Alignment.}  
Across most features, the gating weights show strong alignment between predicted and true congestion labels. This validates that the model is not only fitting the training objective, but that the learned preferences generalize to meaningful physical properties in actual designs.

\paragraph{Observation 2: Unique Negative Attribution.}  
Among all 25 features, only one — \texttt{macro\_rudy\_boundary\_interaction\_index} — consistently receives a negative gating weight. This aligns with earlier findings in Section~\ref{sec:analysis:attribution}, where this feature ranked highly (inversely) correlated with congestion. The model correctly assigns it a down-weighted preference, consistent with human-understandable design heuristics.

\paragraph{Observation 3: Gating Weights Capture Context Beyond Feature Values.}  
Notably, the spread of gating weights is often broader than that of feature values. In some features, the same value results in different gating weights across samples. This suggests that the model modulates importance dynamically based on context — likely derived from the input image, configuration tokens, or pretraining of the MLLM backbone. This further supports the idea that gating weights encode richer, more contextualized preference signals than raw features alone.

\begin{figure}[ht]
    \centering
    \includegraphics[width=\linewidth]{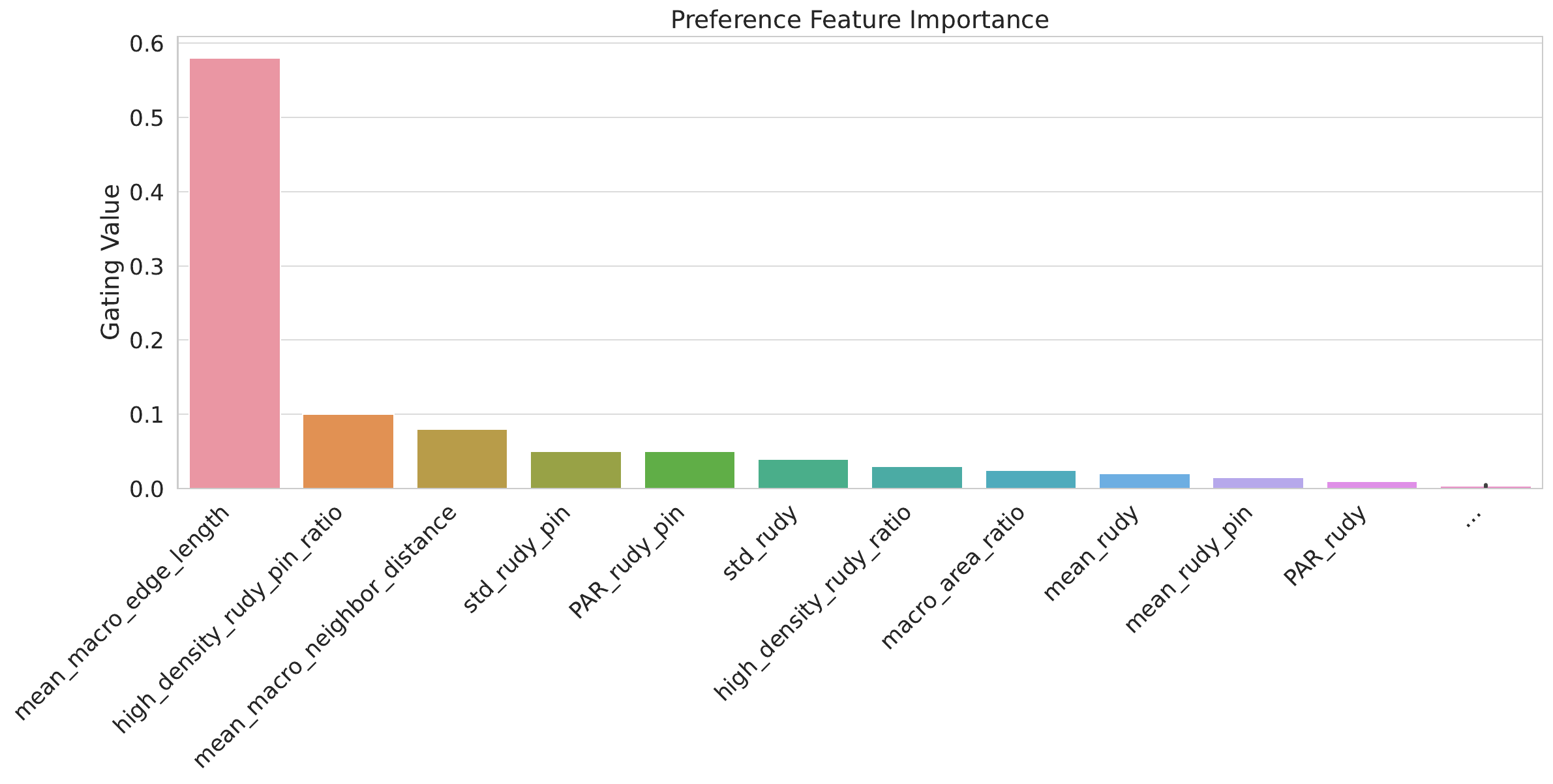}
    \vspace{-0.5cm}
    \caption{
    Bar plot showing the relative importance based on gating values from the preference features used for congestion prediction. 
    The most influential feature is \texttt{mean\_macro\_edge\_length}, significantly outweighing others. 
    This suggests that macro-level spatial layout characteristics play a dominant role in determining congestion levels, followed by pin-density features such as \texttt{high\_density\_rudy\_pin\_ratio} and \texttt{mean\_macro\_neighbor\_distance}.
    }
    \label{app:feature-attribution}
\end{figure}

\clearpage

\section{Detailed Ablation Study}
\label{app:ablation}

\begin{table}[h]
\centering
\caption{Ablation Studies.}
\label{tab:ablation-results}

\begin{subtable}[t]{0.49\textwidth}
    \centering
    \vspace{0pt} % force top alignment
    \caption{Ablation on feature generation pipeline methods.}
    \label{tab:ablation-prompt}
    \resizebox{\linewidth}{!}{%

 \begin{tabular}{l|ccc|ccc}
\toprule
\multirow{2}{*}{\textbf{Pipeline Variant}} & \multicolumn{3}{c|}{\textbf{zero-riscy-a}} & \multicolumn{3}{c}{\textbf{zero-riscy-b}} \\
& \textbf{PLCC ↑} & \textbf{SRCC ↑} & \textbf{KRCC ↑} & \textbf{PLCC ↑} & \textbf{SRCC ↑} & \textbf{KRCC ↑} \\
\midrule
Crossover-Only                    & 0.630 & 0.674 & 0.492  & 0.589 &\textbf{ 0.686} & \textbf{0.493}\\
Mutation-Only              & 0.353 & 0.485 & 0.348 & 0.469 & 0.462 & 0.314 \\
Genetic Instruct        & \textbf{0.705} & \textbf{0.786} & \textbf{0.595} & \textbf{0.599} & 0.591 & 0.412 \\
\bottomrule
\end{tabular}
    }
\end{subtable}
\hfill
\begin{subtable}[t]{0.49\textwidth}
    \centering
    \vspace{0pt} % force top alignment
    \caption{Ablation on interpretable features.}
    \label{tab:ablation-preference}
    \resizebox{\linewidth}{!}{%
    \begin{tabular}{l|c|c}
\toprule
\textbf{Feature Source Modality} & \textbf{SSIM ↑} & \textbf{NRMSE ↓} \\
\midrule
Image-Based & \textbf{0.791} & 0.047 \\
Log-Based   & 0.764 & 0.052 \\
Mixed (Image + Log)      & 0.789 & \textbf{0.046} \\
\bottomrule
\end{tabular}
    }
\end{subtable}
\hfill
\begin{subtable}[t]{0.49\textwidth}
    \centering
    \vspace{0pt} % force top alignment
    \caption{Ablation on conditional states.}
    \label{tab:ablation-modalities}
    \resizebox{\linewidth}{!}{%
    \begin{tabular}{l|c|c}
\toprule
\textbf{Conditional Modality} & \textbf{SSIM ↑} & \textbf{NRMSE ↓} \\
\midrule
Features and Gating weights & & \\
- numerical format (MLP) & 0.787 & 0.047 \\ 
- text format (T5 Encoder)           & \textbf{0.791} & 0.047 \\
\bottomrule
\end{tabular}
    }
\end{subtable}
\hfill
\begin{subtable}[t]{0.49\textwidth}
    \centering
    \vspace{0pt} % force top alignment
    \caption{Ablation on loss functions.}
    \label{tab:ablation-loss}
    \resizebox{\linewidth}{!}{%
    \begin{tabular}{l|c|c}
        \toprule
        \textbf{Loss Variant} & \textbf{SSIM ↑} & \textbf{NRMSE ↓} \\
        \midrule
        U-Net with text conditioning & & \\
        - MSE & \textbf{0.791} & \textbf{0.047} \\
        - MPGD & 0.789 & \textbf{0.047} \\
        - MS-SSIM + L1 & 0.762 & 0.079 \\
        \bottomrule
    \end{tabular}
    }
\end{subtable}
\hfill
\begin{subtable}[t]{0.49\textwidth}
    \centering
    \vspace{0pt} % force top alignment
    \caption{Ablation on congestion map generation modules.}
    \label{tab:ablation-arch}
    \resizebox{\linewidth}{!}{%
    \begin{tabular}{l|c|c}
        \toprule
        \textbf{Input Variant} & \textbf{SSIM ↑} & \textbf{NRMSE ↓} \\
        \midrule
        U-Net & 0.791 & 0.047 \\
        U-Net + CNN Highway & 0.792 & 0.048 \\
        U-Net + MLLM (trainable) & \textbf{0.807} &\textbf{0.045} \\
        \bottomrule
    \end{tabular}
    }
\end{subtable}
\end{table}

\section{Experiment Setup}
\label{app:setup}

\subsection{Training Setup}
All models are trained using the AdamW optimizer with mixed-precision (FP16) training enabled. Most experiments are conducted on 32 NVIDIA A100 GPUs, while some smaller-scale experiments and ablations are run on NVIDIA RTX A6000 GPUs. We train for 50 epochs with early stopping based on validation loss. The learning rate follows a linear warm-up over the first 500 steps and a cosine decay schedule thereafter. The dataset is split into 80\% training, 10\% validation, and 10\% test. All reported results are based on the held-out test set. To ensure robustness, each experiment is repeated with three random seeds.

\subsection{Hyperparameters}
We set the initial learning rate to 1e-4 with weight decay of 0.01. The optimizer uses $\beta_1=0.9$ and $\beta_2=0.999$. The batch size is set to 4. We adopt MiniCPM-V-2\_6~\citep{yao2024minicpm} as the backbone of our MLLM, where the visual encoder follows a SigLIP-like architecture.  For the regression layer encoder, we use a linear layer appended to the backbone; for the gating layer, we use a ReLU MLP of 3 hidden layers with 1024 hidden
units. The conditioning text descriptions for U-Net are encoded with a frozen T5 encoder model unless otherwise stated. Dropout of 0.1 is applied to all modality-specific encoders.

\subsection{Dataset Preprocessing}
% We preprocess the CircuitNet dataset to produce aligned multimodal representations consisting of:
% \begin{itemize}
%     \item Layout-based raster images (Macro Region, RUDY, RUDY pin)
%     \item Normalized tabular metrics extracted via code-driven feature engineering
%     \item Optional config text describing design modules and targets
% \end{itemize}
We preprocess the CircuitNet dataset to construct aligned multimodal representations, including: (1) layout-based raster images (e.g., Macro Region, RUDY, and RUDY pin maps), (2) normalized tabular features extracted through code-driven analysis, and (3) optional configuration text describing design modules and target specifications.
We partition the dataset into 80\% training, 10\% validation, and 10\% test. All results reported are on held-out test samples. Additional augmentations include rotational invariance for layout images and controlled masking of tabular features to assess robustness.

\clearpage
\section{Complete List of Generated Feature Pool and Feature Engineering Details}
\label{app:feature-pool}

\begin{figure}[h]
\small
\begin{tcolorbox}[title=Image-based Feature Pool (Appendix), colback=white!95!gray, colframe=black!75!black, sharp corners=south]
\tiny
\begin{tabular}{p{0.32\textwidth} p{0.65\textwidth}}
\textbf{Feature Name} & \textbf{Description} \\
\midrule
\texttt{rudy\_gradient\_variability} & the variation in gradient changes across the rudy map indicating potential areas of abrupt routing demand shifts \\ 
\texttt{clustered\_macro\_distance\_std} & the standard deviation of distances between clustered groups of macros \\
\texttt{rudy\_pin\_clustering\_coefficient} & a measure of how many rudy pins cluster together relative to the total number of rudy pins \\
\texttt{macro\_density\_gradient} & the change in macro density across the layout, impacting local congestion \\
\texttt{macro\_aspect\_ratio\_variance} & the variance in aspect ratios of macros, indicating potential alignment and spacing issues that may impact congestion \\
\texttt{macro\_compactness\_index} & a measure of how closely packed the macros are, potentially affecting routing paths and congestion \\
\texttt{rudy\_pin\_compaction\_ratio} & the ratio of compacted rudy pin clusters to the total number of rudy pins, indicating areas with high potential routing conflicts \\
\texttt{macro\_variability\_coefficient} & a measure of the consistency in macro sizes and shapes relative to each other, potentially affecting congestion balance \\
\texttt{macro\_cluster\_density\_contrast} & the contrast in density between clustered groups of macros and their surrounding layout areas, indicating potential localized congestion pressure \\
\texttt{rudy\_pin\_distribution\_kurtosis} & a measure of the peakedness or flatness in the distribution of rudy pins across the layout \\
\texttt{localized\_rudy\_variability\_coefficient} & variation in RUDY intensity within localized regions, indicating potential micro-level congestion fluctuations \\
\texttt{macro\_distribution\_clarity\_index} & a measure of how distinct macro distributions are across the layout, indicating clarity in separation \\
\texttt{rudy\_direction\_consistency\_index} & a measure of the uniformity in the directional flow of RUDY intensity \\
\texttt{rudy\_pin\_area\_masking\_index} & the ratio of the area masked by rudy pin regions relative to the total layout \\
\texttt{rudy\_pin\_gradient\_convergence} & a measure of how gradients in the rudy pin map converge into specific regions \\
\texttt{rudy\_deviation\_effect\_index} & the deviation of RUDY intensities from the mean, indicating areas of abnormal routing demand \\
\texttt{demarcated\_macro\_proximity\_index} & proximity of macros to predefined boundary regions, potentially affecting congestion near edges \\
\texttt{macro\_surface\_irregularity\_index} & the irregularity in macro surface shapes, which can impact routing paths \\
\texttt{macro\_rudy\_boundary\_interaction\_index} & interaction between macros and high RUDY regions, indicating potential congestion hotspots \\
\texttt{pin\_density\_peak\_contrast} & the contrast between peak pin density regions and their surroundings \\
\texttt{rudy\_pin\_density\_flux\_index} & the rate of change in rudy pin density across the layout \\
\texttt{high\_density\_rudy\_ratio} & the ratio of areas with high RUDY intensity to the total layout area \\
\texttt{high\_density\_rudy\_pin\_ratio} & the ratio of areas with high RUDY pin intensity to the total layout area \\
\end{tabular}
\end{tcolorbox}
\caption{Complete list of generated features used in our modeling pipeline. These capture spatial, statistical, and gradient-based properties derived from chip layout images.}
\end{figure}

\begin{figure}[h]
\begin{tcolorbox}[title=Configuration and Log-derived Feature Pool, colback=white!95!gray, colframe=black!75!black, sharp corners=south]
\scriptsize
\begin{tabular}{p{0.32\textwidth} p{0.65\textwidth}}
\textbf{Feature Name} & \textbf{Description} \\
\midrule
\texttt{design\_name} & Design identifier (e.g., \texttt{RISCY-a}), useful for tracking design variants. \\
\texttt{number\_of\_macros} & Total number of macros in the design, affecting placement complexity. \\
\texttt{clock\_frequency} & Operating frequency (e.g., 200MHz), indicating timing pressure that may affect placement/routing. \\
\texttt{utilization} & Target utilization (e.g., 70\%), influencing available routing resources. \\
\texttt{macro\_placement} & Macro placement strategy or group count, impacting layout structure. \\
\texttt{power\_mesh\_setting} & Power mesh granularity (e.g., 8), which may influence routing congestion and IR-drop safety margins. \\
\texttt{filler\_insertion} & Stage of filler insertion (e.g., after placement), potentially affecting congestion distribution. \\
\texttt{initial\_placement\_efficiency} & Efficiency of macro and cell placement before routing begins. \\
\texttt{instance\_blockages\_count} & Number of instance blockages, reflecting routing resource obstruction from macro/cell placement. \\
\texttt{hard\_to\_access\_pins\_ratio} & Proportion of pins difficult to route due to placement or geometry constraints. \\
\texttt{pin\_density\_variance\_map} & Variance in pin density across layout regions, highlighting routing hotspots. \\
\texttt{multi\_layer\_pin\_access\_variability} & Variation in pin accessibility across metal layers, indicating routing complexity. \\
\texttt{non\_default\_routing\_rule\_usage} & Frequency of non-default routing rules (if known from constraints or synthesis config). \\
\texttt{crosstalk\_sensitive\_zones} & Areas near critical nets where routing rules must prevent signal interference. \\
\end{tabular}
\end{tcolorbox}
\caption{Complete list of configuration- and log-derived features used in our modeling pipeline. These include routing and congestion-related metrics obtained from the global placement and routing stages, as well as features generated by our own pipeline. They collectively capture layout-level difficulty indicators relevant for congestion prediction.}
\end{figure}

\begin{figure}[h]
\centering
\begin{tcolorbox}[title=Feature Extraction Code Example: Macro Density Gradient, colback=white!95!gray, colframe=black!75!black, sharp corners=south]
\begin{lstlisting}[language=Python, basicstyle=\scriptsize\ttfamily, breaklines=true]
def macro_density_gradient(images):
    tiles_size = 2.25
    macro_image = images[0]
    rudy_image = images[1]
    rudy_pin_image = images[2]
    
    image_height, image_width = macro_image.shape
    total_image_area = image_width * image_height
    
    # Convert macro image to binary [0, 255]
    macro_image = np.uint8(macro_image * 255)
    _, binary_image = cv2.threshold(macro_image, 127, 255, cv2.THRESH_BINARY)
    
    # Find contours to get macro regions
    contours, _ = cv2.findContours(binary_image, cv2.RETR_EXTERNAL, cv2.CHAIN_APPROX_SIMPLE)
    
    # Calculate macro density per region
    macro_density = np.zeros((image_height, image_width))
    for contour in contours:
        mask = np.zeros_like(binary_image)
        cv2.drawContours(mask, [contour], -1, 255, thickness=cv2.FILLED)
        macro_density += mask

    # Gradient of the macro density
    gradient_x = cv2.Sobel(macro_density, cv2.CV_64F, 1, 0, ksize=5)
    gradient_y = cv2.Sobel(macro_density, cv2.CV_64F, 0, 1, ksize=5)
    
    # Calculate the magnitude of gradients
    gradient_magnitude = cv2.magnitude(gradient_x, gradient_y)

    # Calculate the average gradient magnitude in micrometers
    macro_density_gradient_um = np.sum(gradient_magnitude) / (image_height * image_width)
    macro_density_gradient_um *= tiles_size  # Convert to micrometers

    return {"macro_density_gradient": macro_density_gradient_um}
\end{lstlisting}
\end{tcolorbox}
\caption{Python implementation of the \texttt{macro\_density\_gradient} feature extractor. This function computes the average gradient magnitude of macro cell density across a chip layout image, offering a scalar representation of spatial density variation in micrometers.}
\end{figure}

\begin{figure}[h]
\centering
\begin{tcolorbox}[title=Feature Extraction Code Example: High Density Rudy Ratio, colback=white!95!gray, colframe=black!75!black, sharp corners=south]
\begin{lstlisting}[language=Python, basicstyle=\scriptsize\ttfamily, breaklines=true]
def high_density_rudy_ratio(images):
    image = images[1]
    total_area = image.shape[0] * image.shape[1]
    mean_rudy = np.mean(image)
    high_density_rudy_ratio = (image > mean_rudy).sum() /  total_area
    
    return {
        "high_density_rudy_ratio": high_density_rudy_ratio,
    }
\end{lstlisting}
\end{tcolorbox}
\caption{Python implementation of the \texttt{high\_density\_rudy\_ratio} feature extractor. This function computes the average gradient magnitude of macro cell density across a chip layout image, offering a scalar representation of spatial density variation in micrometers.}
\end{figure}

\end{document}